\documentclass[12pt,superscriptaddress,showpacs]{iopart}
\usepackage{graphicx,iopams}
\usepackage{bm}
\input epsf.sty

\begin{document}

\title{Persistent currents in a graphene ring with armchair edges}
\author{Bor-Luen Huang}
\address{Department of Physics, National Taiwan Normal University, Taipei 11677, Taiwan}
\author{Ming-Che Chang}
\address{Department of Physics, National Taiwan Normal University, Taipei 11677, Taiwan}
\author{Chung-Yu Mou}
\address{Department of Physics, National Tsing Hua University, Hsinchu 30013,
Taiwan}\address{Institute of Physics, Academia Sinica, Nankang 11529, Taiwan} \address{Physics
Division, National Center for Theoretical Sciences, P.O.Box 2-131, Hsinchu, Taiwan}

\begin{abstract}
A graphene nano-ribbon with armchair edges is known to have no edge state. However, if the
nano-ribbon is in the quantum spin Hall state, then there must be helical edge states. By
folding a graphene ribbon to a ring and threading it by a magnetic flux, we study the
persistent charge and spin currents in the tight-binding limit. It is found that, for a broad
ribbon, the edge spin current approaches a finite value independent of the radius of the ring.
For a narrow ribbon, inter-edge coupling between the edge states could open the Dirac gap and
reduce the overall persistent currents. Furthermore, by enhancing the Rashba coupling, we find
that the persistent spin current gradually reduces to zero at a critical value, beyond which
the graphene is no longer a quantum spin Hall insulator.
\end{abstract}

\maketitle

\section{Introduction}

A crucial feature of topological insulator is the existence of metallic surface states
\cite{Hasan10}. These surface states (or {\it edge states} if the insulator is two dimensional)
have odd pairs of channels, in which each pair consists of opposite spins moving in opposite
directions (helical edge states). Therefore, they could carry charge as well as spin currents.
We are interested in studying their transport using a lattice model. In theory, one could rely
on a model with external leads and reservoirs, or with a closed ring in which the currents are
driven by a threading magnetic flux. In this work, we adopt the latter (simpler) approach, and
use the graphene lattice as a theoretical model to study the transport of these important
helical edge states.

Graphene is one of the earliest candidates of two-dimensional topological insulator (also
called quantum spin Hall insulator) \cite{Kane05}. It is found that, graphene would become a
quantum spin Hall (QSH) insulator if there is a spin-orbit interaction (SOI). However, since
graphene's SOI is only of the order of $10^{-3}$ meV \cite{Min06}, it poses an experimental
challenge to probe such a state \cite{Weeks11}. Experimental issues aside, graphene lattice is
ideal for theoretical investigation of the helical edge states. To avoid complications from
pre-existing edge states, we choose a graphene ribbon with armchair edges, which has no edge
state in the absence of SOI. That is, the edge states studied here are intrinsic to the QSH
insulator. In contrast, a graphene ribbon with zigzag edges has edge states even without SOI
\cite{Nakada96}. These edge states would become helical edge states (similar to those in this
work) with SOI.

If the graphene ribbon is wrapped up to form a tube (or a ring, with armchair edges), with a
magnetic flux passing through the hole, then the helical edge state provides a robust channel
for persistent charge current (PCC) and persistent spin current (PSC). PCC in a mesoscopic
metal ring was first predicted \cite{Buttiker}, and later verified in
experiments\cite{Chandrasekhar} decades ago. In a recent work \cite{Jayich09}, the PCC was
measured accurately by mounting nano-rings on a micro-cantilever in an external magnetic field.
In addition to PCC, the PSC in a ring in a textured magnetic field \cite{Loss}, or in a ring
with SOI \cite{Splettstoesser03} has been proposed. Further researchers show that it is
possible to have a PSC in a ferromagnetic ring \cite{Kollar01} or an antiferromagnetic ring
\cite{Kollar02}.

There have been several studies of the Aharonov-Bohm effect and PCC in a graphene ring with
different geometrical shapes (without spin-orbit interaction). For example, a nano-torus
\cite{Lin98}, flat rings (similar to Corbino-disks) with various types of boundaries
\cite{Recher07,Schelter10,Ma10}, or a tube with zigzag edges \cite{Dutta11}. Observation of the
Aharonov-Bohm effect is an indirect evidence of the PCC. Recently, the Aharonov-Bohm
oscillation in a nano-rod of topological insulator has been observed \cite{Peng10} and studied
\cite{Bardarson10,Zhang10}. Theoretical investigations of the persistent currents in a
two-dimensional topological-insulator ring, which are based on an effective continuous model,
can be found in Refs.~\cite{Michetti11,Chang10}. In Ref.~\cite{Soriano10}, the zigzag edge of a
graphene sheet is found to be magnetized due to the Hubbard interaction. Since the time
reversal symmetry is spontaneously broken, edge charge current appears without the need of a
driving magnetic flux.

In this paper, we study the quantum spin-Hall states and the persistent currents of an armchair
ring with SOI. Different from earlier approaches \cite{Michetti11,Chang10}, this work is based
on a lattice model, which allows insulating band structure and genuine edge states. By changing
its parameters, the gaphene ring can be in or out of the QSH phase, and the behavior of the
persistent currents across the phase boundary can be studied. It is found that: For a broad
ribbon, the edge PCC is small (but non-zero), while the edge PSC approaches a finite value. For
a narrow ribbon, the edge states from opposite sides couple with each other, which may open an
energy gap and reduce the persistent currents. We have also considered the Zeeman coupling and
the Rashba coupling. The former introduces addition jumps in the periodic saw-tooth curves of
the persistent currents. By breaking spin conservation, the Rashba coupling reduces the edge
spin current, which eventually vanishes at the boundary of the QSH phase. Many of the results
obtained here should also apply qualitatively to the helical edge states and persistent
currents of a three-dimensional topological-insulator ring.

The paper is outlined as follows: In Sec.~2, we describe the theoretical model being used. In
Sec.~3 we first study the helical edge states and related PCC and PSC in rings with various
sizes, then comment on the effects of the Zeeman and the Rashba interactions. Sec.~4 is the
conclusion.

\section{Theoretical formulation}

\begin{figure}
\center
\includegraphics*[width=7cm]{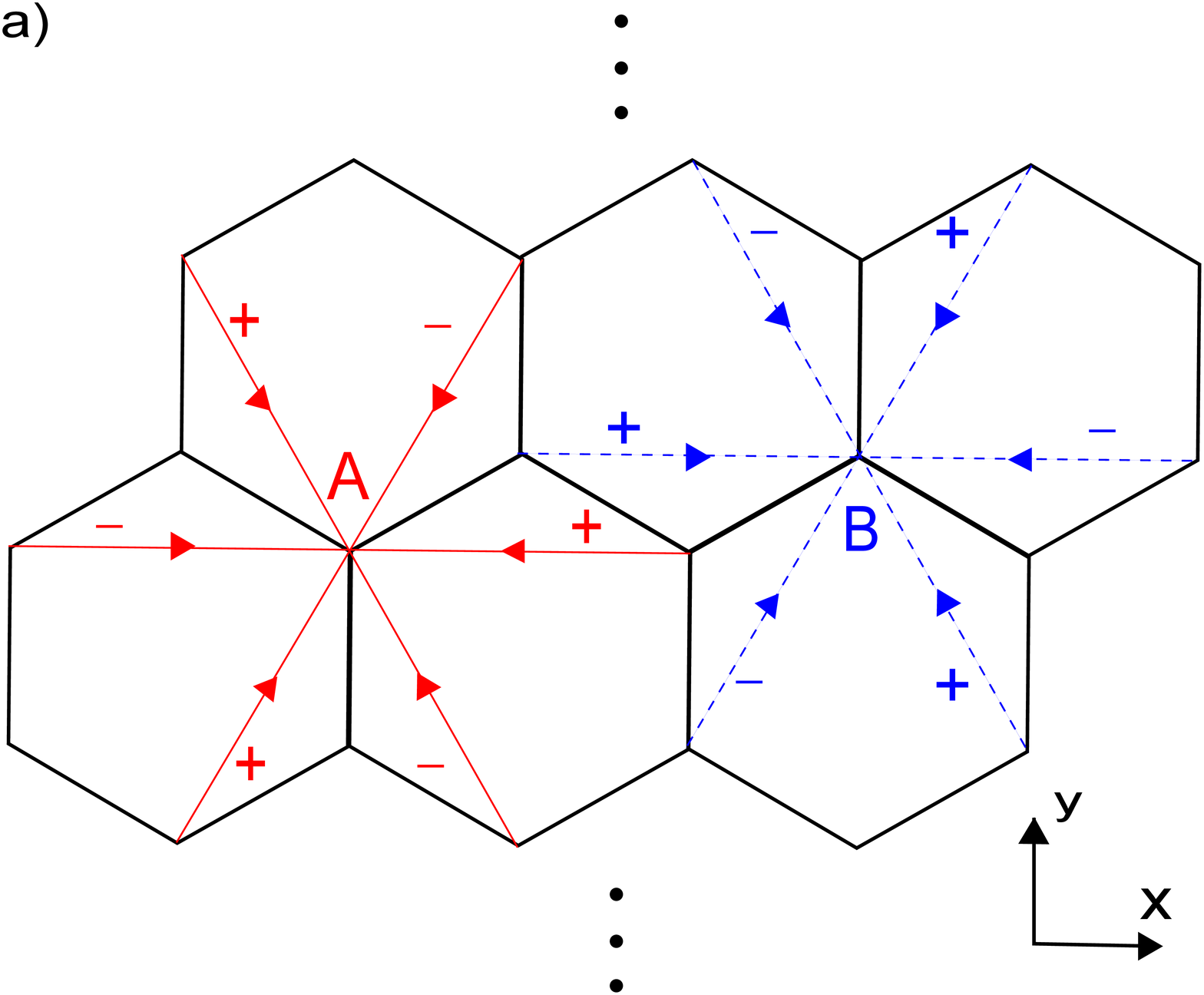}\vspace{3mm}
\center
\includegraphics*[width=7.2cm]{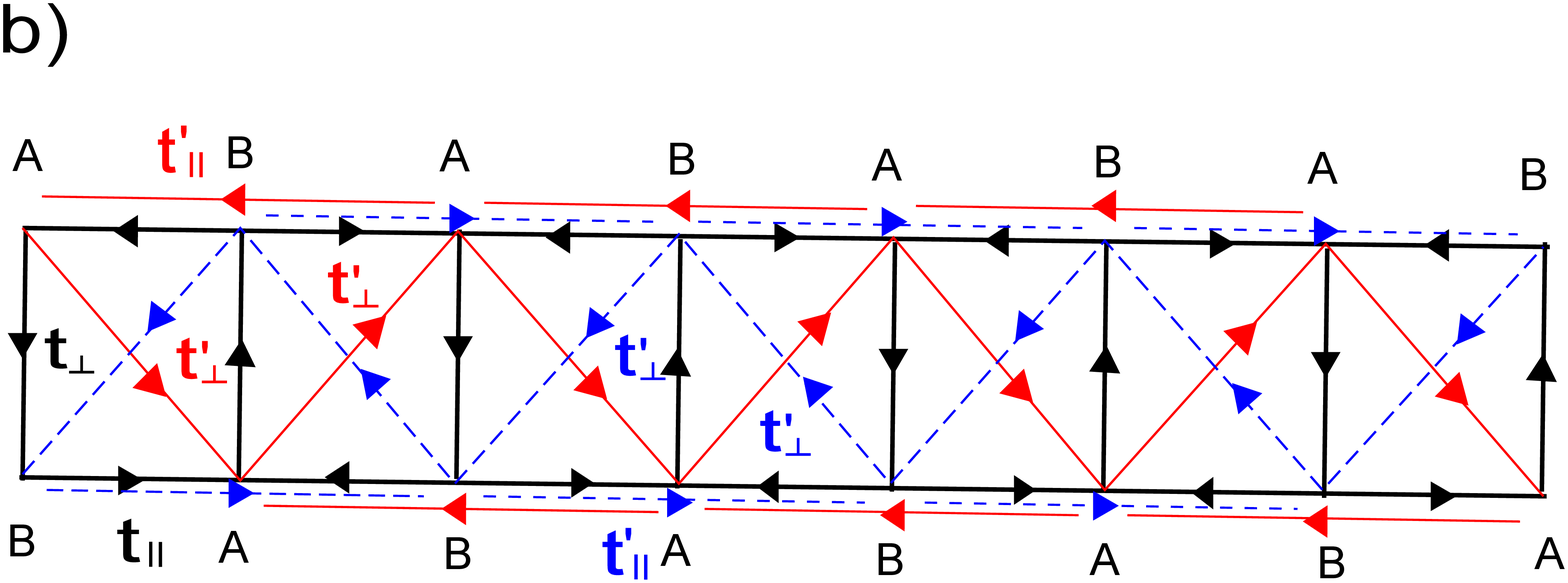}\vspace{3mm}
\caption{The honeycomb lattice is shown in (a). Red and blue arrows are the
next-nearest-neighbor hoppings. The red and blue arrows in the plot are for spin-up electrons
in $A$- and $B$-sublattices, and $+$ and $-$ signs refer to the signs of $\nu_{ij}$. For
spin-down electrons, red and blue arrows appear in opposite directions. The effective two-lag
ladder for the Hamiltonian in Eq.~(3) is shown in (b).}\label{lattice}
\end{figure}

We start with a tight-binding model for a graphene ribbon. One imposes the open boundary
condition with armchair edges along the $x$-direction, and periodic boundary condition along
the $y$-direction (see Fig.~\ref{lattice}). A magnetic flux $\Phi$ passes through the inside of
the ring along the $(-x)$-direction. Because of the magnetic flux, an electron circling the
ribbon once would acquire an Aharonov-Bohm phase of $2\pi\Phi/\Phi_0$, where $\Phi_0$ is the
flux quantum. The Hamiltonian reads \cite{Kane05},
\begin{eqnarray}
H&=&-t\sum_{<i,j>}e^{i\theta_1}c^{\dag}_ic_j+i\lambda_{SO}\sum_{<<i,j>>}e^{i\theta_2}\nu_{ij}
c^{\dag}_i \sigma^z c_j\\ \nonumber &+&\frac{g}{2}\mu_B B\sum_i c_i^\dagger\sigma^x c_i
+i\lambda_R\sum_{<i,j>}e^{i\theta_1}c_i^\dagger(\mbox{{\boldmath$\sigma$}}\times{\hat{\bf
d}}_{ij})_z c_j,
\end{eqnarray}
where $c^\dag_i=(c^\dag_{i\uparrow},c^\dag_{i\downarrow})$ creates an electron at lattice site
$i$. The first term accounts for the nearest-neighbor hoppings.  The second term is a SOI with
next-nearest-neighbor hoppings, in which $\nu_{ij}=(2/\sqrt{3})(\hat{d_1} \times
\hat{d_2})\cdot\hat{z}$, $\hat{d_1}$ and $\hat{d_2}$ are the two nearest-neighbor bonds that
connect site-$j$ to site-$i$, and $\mbox{\boldmath{$\sigma$}}$ are the Pauli matrices. The
third and the fourth terms are the Zeeman and the Rashba couplings. In the following, all
energies and lengths will be in units of $t$ and $a$ (lattice constant) respectively.

The phases $\theta_1$ and $\theta_2$ are related to the magnetic flux: $\theta_1=2\phi/3$ for
the nearest-neighbor bonds along the $y$-direction (see Fig.~\ref{lattice}), while
$\theta_1=\phi/3$ for the zigzag bonds, where $\phi=(2\pi/N_y)\Phi/\Phi_0$, $N_y$ is the number
of unit cells around the ring. The phase $\theta_2$ is zero for the next-nearest-neighbor bonds
along the $x$-direction, while $\theta_2=\phi$ for the other next-nearest-neighbor bonds. These
phases are chosen in such a way that the electron would pick up the correct Aharonov-Bohm phase
when circling the ring once, no matter which path it takes. The geometric curvature of the tube
is known to enhance the SOI \cite{Huertas06}, but such a curvature effect is not considered.

Because of the translational invariance along the $y$-direction, one can perform the following
Fourier transformation
\begin{equation}
c_{x,y}=\frac{1}{\sqrt{N_y}}\sum_{k_y}e^{i k_y y}c_{x,k_y}
\end{equation}
and map the system to a two-lag ladder \cite{Wu}. The effective Hamiltonian can be written as,
\begin{equation}
H=\sum_{k_y}\Psi^\dag\left(
\begin{array}{cccc}
T' & T & \mu_B B & R \\
T^\dagger & -T' & R^* & \mu_B B \\
\mu_B B & R^T & -T' & T \\
R^\dagger & \mu_B B & T^\dagger & T'
\end{array}
\right)\Psi,
\end{equation}
where $\Psi^\dag(x,k_y)=(\psi^\dag_{A\uparrow}, \psi^\dag_{B\uparrow}, \psi^\dag_{A\downarrow},
\psi^\dag_{B\downarrow})$. The matrix elements are
\begin{equation}
T'=\left(
\begin{array}{cccccc}
0 & t'^{*}_\perp & t'_\parallel & 0 & \cdots & 0 \\
t'_\perp & 0 & t'^{*}_\perp & t'_\parallel & \ddots & \vdots \\
t'^{*}_\parallel & t'_\perp & 0 & t'^{*}_\perp & \ddots & 0 \\
0 & t'^{*}_\parallel & t'_\perp & 0 & \ddots & t'_\parallel \\
\vdots & \ddots & \ddots & \ddots & \ddots & t'^{*}_\perp \\
0 & \cdots & 0 & t'^{*}_\parallel & t'_\perp & 0
\end{array}
\right),
\end{equation}
where $ t_\parallel=te^{i\frac{1}{3}\phi_k}$, $t_\perp=te^{i\frac{2}{3}\phi_k}$,
$t'_\parallel=i\lambda_{SO}$, and $ t'_\perp=2i\lambda_{SO}\cos{\phi_k}$, where
$\phi_k=\frac{\sqrt{3}}{2}k_y a-\phi/2$, and
\begin{equation}
T=\left(
\begin{array}{ccccc}
t^{*}_\perp & t_\parallel & 0 & \cdots & 0 \\
t_\parallel & t^{*}_\perp & t_\parallel & \ddots & \vdots \\
0 & t_\parallel & t^{*}_\perp & \ddots & 0  \\
\vdots & \ddots & \ddots  & \ddots & t_\parallel \\
0 & \cdots & 0 & t_\parallel & t^{*}_\perp
\end{array}
\right),
\end{equation}
In Eq.~(3), $\mu_B B$ is multiplied by an unit matrix. The Rashba term gives
\begin{equation}
R=\left(
\begin{array}{ccccc}
R_\perp^* & R_- & 0 & \cdots & 0 \\
R_+ & R_\perp^* & R_-  & \ddots & \vdots \\
0 & R_+ & R_\perp^*  & \ddots & 0 \\
\vdots & \ddots & \ddots  & \ddots & R_- \\
0 & \cdots & 0  & R_+ & R_\perp^*
\end{array}
\right),
\end{equation}
where $R_\perp=-\frac{i}{\sqrt{3}}\lambda_R e^{2i\phi_k/3}$, $R_\pm=R_y\pm iR_x$,
$R_x=\frac{i}{2}\lambda_R e^{i\phi_k/3}$, and $R_y=-\frac{i}{2\sqrt{3}}\lambda_R
e^{i\phi_k/3}$.

At zero temperature, the PCC of the nano-ring is given by (in units of $et/h$),
\begin{equation}
I=-\sum_{m k_y}\frac{\partial E_{m k_y}}{\partial \Phi},\label{PCC}
\end{equation}
in which one sums over filled states (for total current) or edge states (for edge current). The
PSC for spin-$z$ will be calculated by the following semiclassical expression
\cite{Splettstoesser03} (in units of $(\hbar/2)t/h=t/4\pi$)
\begin{equation}
I^s=-\sum_{m k_y}\frac{\partial E_{m k_y}}{\partial \Phi}\langle
mk_y|\sigma_z|mk_y\rangle,\label{semi}
\end{equation}
where $|mk_y\rangle$ are the Bloch states.

When the spin is conserved (Secs.~3.1 and 3.2), the following notations are used: The current
for spin-up and spin-down electrons are $I_\sigma$. Therefore, the total charge current is
$I=I_++I_-$, and the total spin current is $I^s=I_+-I_-$. The spin-dependent edge currents are
$I_{R\sigma}$ and $I_{L\sigma}$. For example, the charge and spin currents for the right edge
are $I_R=I_{R+}+I_{R-}$ and $I^s_R=I_{R+}-I_{R-}$ respectively.

In general, $I_\sigma$ is the sum of edge current and bulk current. The existence of persistent
{\it bulk} current in an insulator ring (the usual type or the QSH type) is due to the
discreteness (due to the finite circumference) and asymmetry (due to the magnetic flux) of the
energy spectrum. As a result, it only exists in a ring with finite radius (with or without
SOI), and diminishes when the ring is larger. On the contrary, there is no spin-resolved
current (either bulk or edge) in an usual insulator ring. It only exists in the QSH phase and
is not reduced in a larger ring. That is, the persistent spin current reported below is
intrinsic to the QSH phase.

\section{Numerical Results}

\begin{figure}
\center
\includegraphics*[width=8cm]{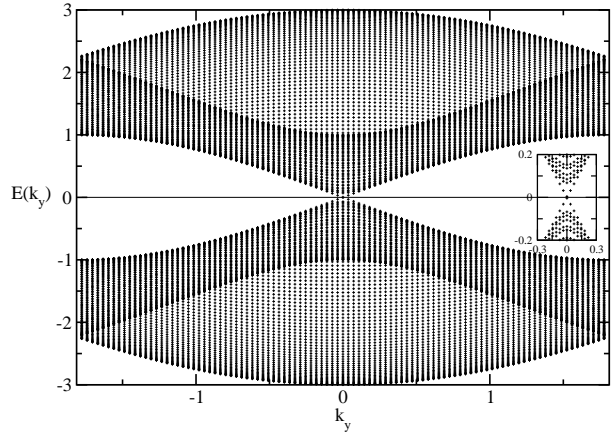}
\caption{Energy spectrum for a graphene ribbon with $n_x=100$, $n_y=100$, and
$\lambda_{SO}=0.01$. The inset shows the magnified view near the Dirac point. The bulk gap is
about $0.136$, which is slightly larger than the bulk value of $6\sqrt{3}\lambda_{SO}$ because
of the finite-size effet. The crossing edge states inside the gap is the main characteristic of
the quantum spin Hall phase.}\label{energy_sp01}
\end{figure}

There are several tunable parameters in the numerical calculations: width and length
(circumference) of the ribbon, spin-orbit couplings $\lambda_{SO}$ and $\lambda_R$, and the
$g$-factor (for the Zeeman coupling). In addition, the magnetic flux inside the ring controls
the magnitude of the persistent currents. To explore such a large parameter space, the
following path is adopted: we first show the energy spectrum and the decay length of the edge
states, then present the charge and spin transport for broad ribbons (Sec.~3.1). Subsequently,
the transport in narrow ribbons are shown and the finite-size effect discussed (Sec.~3.2). In
either case, to simplify the discussion, Zeeman and Rashba couplings are not included. Their
effect will be investigated in Sec.~3.3 and Sec.~3.4 respectively. Among these parameters, the
most important ones are $\lambda_{SO}$ and $\lambda_R$, which dictate whether the graphene is
in the QSH phase \cite{Kane05}. Their overall influence on edge states and spin current can be
seen in Figs. 3 and 10.

Before showing the result, we review the energy spectrum of a {\it flat} graphene nano-ribbon
with armchair edges, but without SOI \cite{Nakada96}. First, the width of the ribbon has a
major effect on the energy gap. If $n_x=3n+2$, then there is a Dirac point at $k_y=0$ and the
system is gapless in the tight-binding model. If $n_x\neq 3n+2$, then the system has a finite
gap, which approaches zero as $n_x$ becomes infinite. Second, in the presence of SOI, the Dirac
point of a ribbon with $n_x=3n+2$ is opened, and mid-gap edge states are crossing at $k_y=0$.
Similar mid-gap edge states exist for $n_x\neq 3n+2$.

\subsection{Broad ribbon}

\begin{figure}
\center
\includegraphics*[width=7cm]{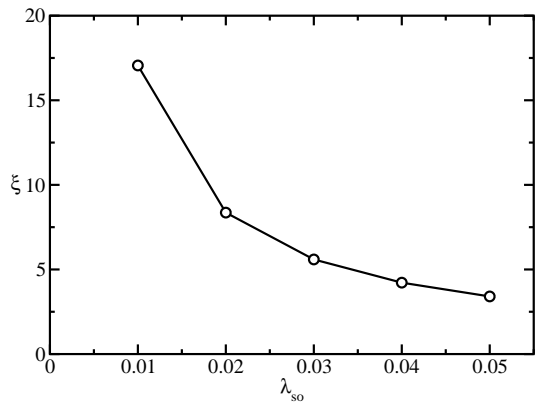}
\caption{Decay length of the edge state at the Dirac point. The size of the ring is $n_x=100$
and $n_y=100$. The decay length is inversely proportional to the strength of the spin-orbit
interaction.}\label{decay_length}
\end{figure}

In Kane and Mele's analysis \cite{Kane05}, the graphene enters the QSH state once the SOI is
turned on, without a finite threshold value. Therefore, the edge modes for the QSH state should
immediately appear, no matter how small the SOI is. In Fig.~\ref{energy_sp01}, we show the
energy dispersion for $\lambda_{SO}=0.01$ and $\Phi=0$. Overall, the energy spectrum shows
intangible change from the one with $\lambda_{SO}=0$. However,  in the inset one can see that
the Dirac point is opened up, with mid-gap edge modes. The edge modes do not merge into bulk
bands up to near the boundaries of the Brillouin zone.

The envelope of the probability distribution for the edge state decays exponentially,
$|\Psi(x)|^2\propto \exp(-d/\xi)$, where $d$ is the distance from the edge. As one increases
the SOI, the edge states should be more localized. In Fig.~\ref{decay_length}, we show the
decay length $\xi$ of the edge state at the Dirac point for different values of
$\lambda_{SO}$. Upon fitting, the inverse of the decay length is about $5.89 \lambda_{SO}$,
proportional to the SOI. For example, for $\lambda_{SO}=0.01 (0.05)$, the decay length is about
17 sites (3 sites). With this result, one can estimate whether, for a given ribbon, the
inter-edge coupling could be safely neglected .

\begin{figure}
\center \rotatebox{0}{\includegraphics*[width=10 cm]{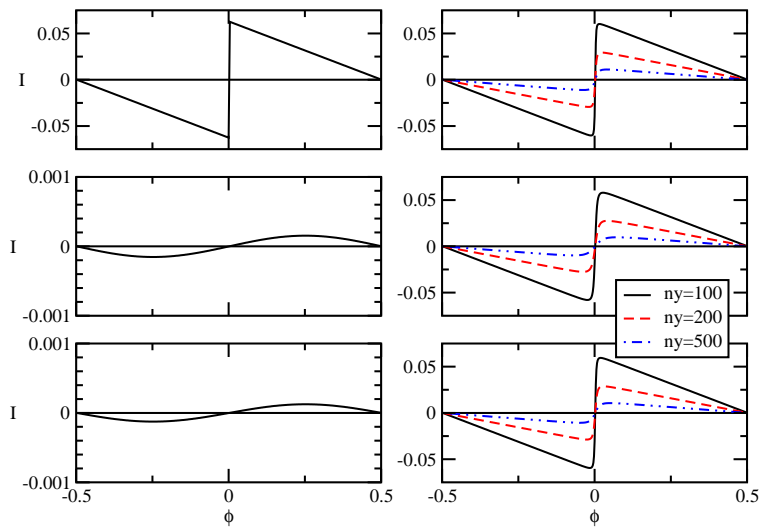}} \caption{The plots on the left
column are the persistent charge currents for $\lambda_{SO}=0$, and those on the right column
are for $\lambda_{SO}=0.05$. The charge currents are in units of $et/h$ (see Eq.\ref{PCC}).The
figures on the first, the second, and the third rows are for $n_x=50, 51$, and $52$
respectively. On the right column, one can see that the persistent currents decrease as $n_y$
becomes larger. The magnitudes for the 3 plots on the right are roughly the same, in sharp
contrast to the ones on the left. } \label{pc_tot}
\end{figure}

At first, we report the PCC {\it without} SOI. In this case, the current is from bulk states
only since there is no edge mode. The PCCs are shown on the left of Fig.~\ref{pc_tot}. As we
mentioned earlier, there is a Dirac point in the energy spectrum when $n_x=3n+2$. The
discontinuity of the slopes at the Dirac point produces the finite jump at zero flux for
$n_x=50$. On the other hand, systems with $n_x\neq 3n+2$ are gapped (without the Dirac point),
so the PCCs show no jump for $n_x=51,52$. Furthermore, the overall magnitude of the PCC is also
much smaller.

The plots for the PCCs {\it with} SOI are shown on the right of Fig.~\ref{pc_tot}. Even though
the SOI introduces little change to the PCC for $n_x=50$, it has a dramatic effect to the
persistent currents of $n_x=51, 52$. In the second and third rows of Fig.~\ref{pc_tot}, the
sinusoidal-like variations on the left become sawtooth-like on the right, and have much larger
magnitudes (similar to the system with $n_x=50$). This, of course, is due to the energy level
crossing of the newly formed edge-modes.

\begin{figure}
\center
\includegraphics*[width=11cm]{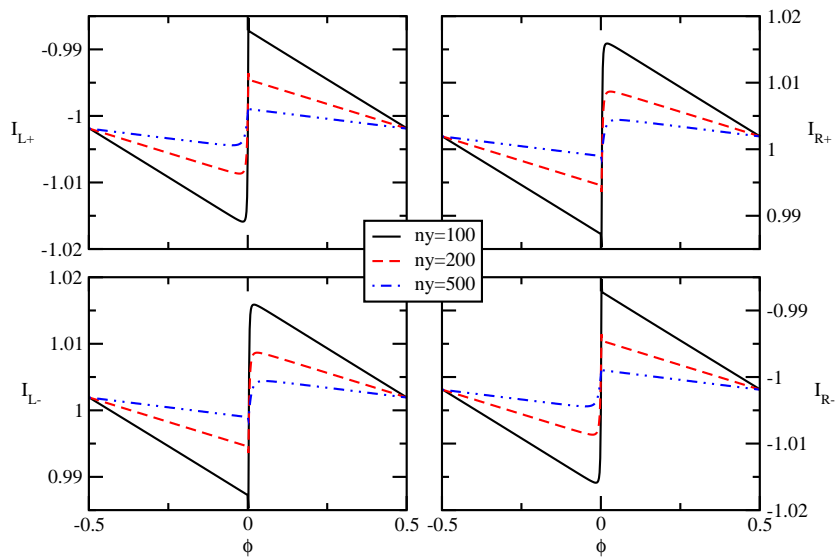}
\caption{Edge currents for spin-up and spin-down electrons for $n_x=50$ and
$\lambda_{SO}=0.05$. When $n_y$ is large, the edge persistent current has a smaller
discontinuity but the overall magnitude remains close to one.} \label{pc_edge}
\end{figure}

We now focus on the edge {\it spin} current in the presence of SOI. Spin-resolved edge currents
$I_{R\sigma}$ ($I_{L\sigma}$) are shown on the right (left) side of Fig.~\ref{pc_edge}. The
spin-resolved edge current has a discontinuity at $\Phi=0$, similar to those in
Fig.~\ref{pc_tot}. However, in Fig.~\ref{pc_edge}, the sawtooth curves ride on background
values of $\pm 1$ (i.e. opposite spins move along opposite directions). Therefore, it is a more
stable transport with respect to the change of magnetic flux. As $n_y$ gets larger, even though
the jump is smaller, the overall magnitudes of these currents remain close to one.

A few symmetries can be observed in Fig.~\ref{pc_edge}. For example, $I_{L+}=I_{R-}$, and
$I_{L-}=I_{R+}$. Also, $I_{R+}(\phi)=-I_{R-}(-\phi)$, similarly for the left edge. Notice that
the charge current for the right edge, $I_R=I_{R+}+I_{R-}$, is small but nonzero due to the
slight asymmetry between $I_{R\pm}(-\phi)$ and $I_{R\pm}(\phi)$. On the other hand, the spin
current for the right edge, $I^s_R=I_{R+}-I_{R-}$, is roughly of value $2$ (in units of
$t/4\pi$).

In the figure, the persistent spin current is nonzero even if there is no magnetic flux. Such a
current, unlike persistent charge current, does not require the breaking of time-reversal
symmetry. However, the edge states being helical is essential. Furthermore, since the magnitude
of the edge spin current is not reduced in a larger ring (in contrast to PCC), it is possible
to have it in a large ring (if phase coherence around the ring could be preserved). For a flat
sheet with open boundary (and no external bias), the edge spin current cannot persist. A
continuous spin circulation without external driving force is possible only in a ring-like
structure. In principle, the moving spins would induce an electric dipole \cite{Kollar01}
(might be oscillating in time) that can be felt. In practice, detecting such an induced
electric field remains a great experimental challenge.

\subsection{Narrow ribbon}

\begin{figure}
\center
\includegraphics*[width=9cm]{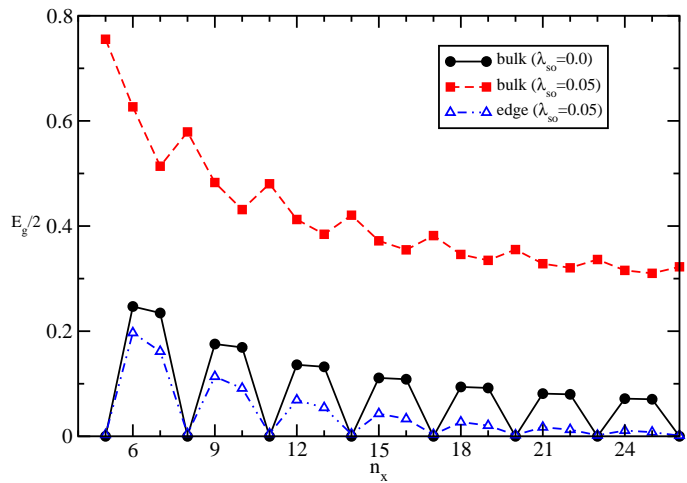}
\caption{Energy gaps for narrow ribbons with $n_y=100$. Solid circles are the energy gaps for
$\lambda_{SO}=0$. The system is semi-metallic for $n_x=3n+2$, and is insulating otherwise. When
$\lambda_{SO}=0.05$, we show both the edge gaps (triangles) and the bulk gaps (solid
squares).}\label{energy_gap}
\end{figure}

As the ribbon becomes narrower, the edge states on the two sides will couple with each other
\cite{Shan10}. Such a coupling may open the Dirac point (when $n_x\neq 3n+2$) at the Fermi
energy. Fig.~\ref{energy_gap} shows the oscillatory energy gaps for different widths of the
ribbon. One can see that the system remains semi-metallic when $n_x=3n+2$, with or without SOI
\cite{Son06}. When $n_x\neq 3n+2$, as $n_x$ becomes smaller, inter-edge coupling produces a
larger gap (triangles in Fig.~\ref{energy_gap}). For reference, the bulk gap at $k_y=0$ is also
plotted (solid squares). Notice that the bulk gap is larger at $n_x= 3n+2$, when the energy gap
for the edge state vanishes.

\begin{figure}
\center
\includegraphics*[width=10cm]{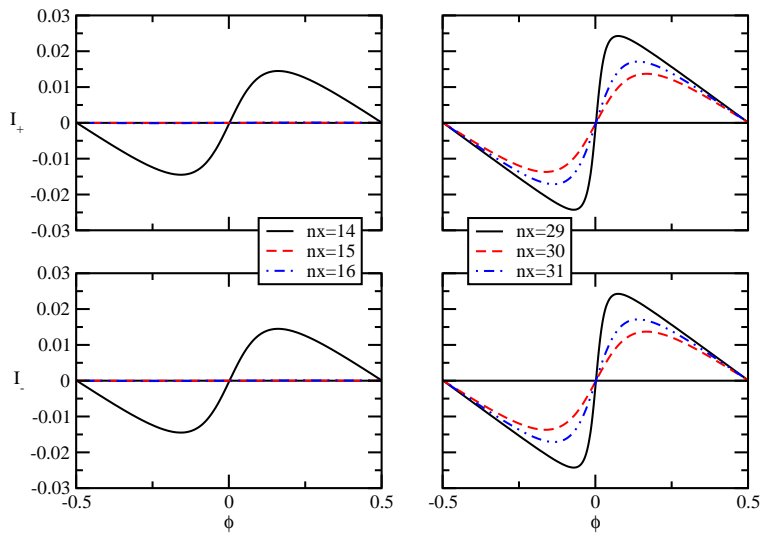}
\caption{Persistent currents for electrons in a narrow ring with $n_y=100$ and
$\lambda_{SO}=0.05$. The left column shows the persistent currents for up and down spins when
$n_x=14,15,16$, while the right column is for $n_x=29,30,31$. }\label{psc_s}
\end{figure}

For a narrow ribbon, the edge modes are no longer well-defined. Therefore, we focus on the
total persistent currents carried by the whole ring. Fig.~\ref{psc_s} shows the persistent
currents of spin-up and spin-down electrons in a narrow ring. As one can see, since $I_+=I_-$,
the charge current $I$ is twice of the values in Fig.~\ref{psc_s}, while the spin current $I^s$
is zero. There could be remnant spin currents along the left and the right edges, but they
would cancel each other. For small widths (the left column), the currents for $n_x\neq 3n+2$
can barely be seen \cite{note1}. This is due to the finite energy gaps shown in
Fig.~\ref{energy_gap}. The case of $n_x= 3n+2$ is special since the Dirac point remains closed.
Such a dramatic contrast is less apparent in the right column.

\subsection{Zeeman interaction}

The analysis so far applies to the cases when the electron in the ring feels the Aharonov-Bohm
effect through the vector potential of the magnetic flux, but not the magnetic field itself. In
the following, we consider a more realistic situation when the graphene ring itself is immersed
in a magnetic field. As a result, the electrons can couple with the field through their spins.
Some order-of-magnitude estimate is in order: For a ring with $n_y$-sites, the cross-sectional
area is $(3/4\pi)n^2_ya^2$ ($a=2.456\AA$). In order to have one flux quantum inside the ring,
the magnetic field needs to be $B\simeq 2.87\times 10^5/n^2_y$ (in units of $T$). The Zeeman
gap at one flux quantum is $33.25/n^2_y$ eV ($g=2$). In comparison, the energy difference
between neighboring $k_y$-states in the Brillouin zone (see Fig.~\ref{energy_sp01}) is roughly
of the order of $2t/n_y$ ($t\simeq 2.8$ eV). Therefore, the Zeeman gap is relatively
unimportant, unless $n_y$ is of the order of ten or less.
\begin{figure}
\center
\includegraphics*[width=9cm]{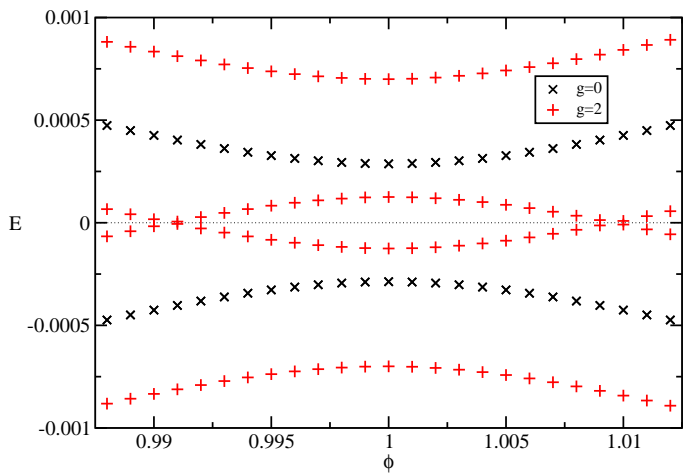}
\caption{The energy levels near $k_y=0$ for a ring with $n_x=50$, $n_y=100$, and
$\lambda_{SO}=0.01$ are plotted with respect to $\phi$ near $\phi=1$. Each curve for $g=0$ is
separated to two curves when $g$ is nonzero.} \label{zeeman1}
\end{figure}

\begin{figure}
\center
\includegraphics*[width=10cm]{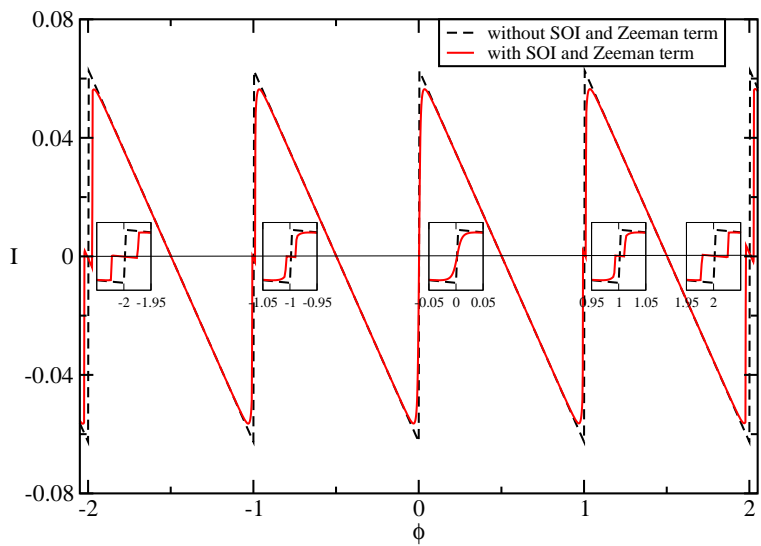}
\caption{Persistent current for a graphene ring with $n_x=50$, $n_y=100$, and
$\lambda_{SO}=0.01$. Insets show magnified views near the jumps.}\label{zeeman2}
\end{figure}

Fig.~\ref{zeeman1} shows the energy levels near the energy gap (with and without the Zeeman
interaction). When $g=0$, there are two parabolas with a minute energy gap due to the
inter-edge coupling. When $g\neq 0$, the upper (and lower) parabola separates into two
parabolas. The separation is proportional to the strength of the Zeeman interaction. As a
result, it is possible for the upper and lower parabolas to intersect with each other and
produce two crossings (two Dirac points). This will alter the qualitative behavior of the
$I(\phi)$ curve.

The Zeeman effect on the PCC can be seen in Fig.~\ref{zeeman2}. Near an integer flux quantum, a
vertical jump splits to two (inset), due to the two Dirac points in Fig.~\ref{zeeman1}. Such a
split is proportional to the magnitude of the magnetic field. As a result, the $I(\phi)$ curve
is no longer periodic in the magnetic flux. This is the main effect of the Zeeman interaction.

\subsection{Rashba interaction}

\begin{figure}
\center
\includegraphics*[width=8cm]{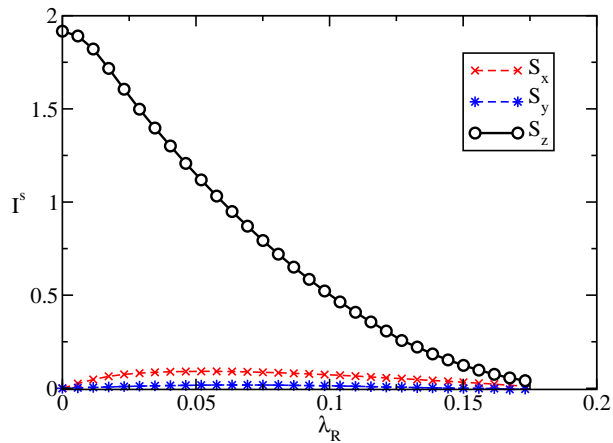}
\caption{Reduction of the edge spin currents (in units of $t/4\pi$) due to the Rashba coupling.
The parameters are $n_x=50, n_y=100$, $\lambda_{SO}=0.05$, and the Zeeman coupling is not
considered.}\label{rashba}
\end{figure}

For a graphene sheet, the Rashba coupling is possible only if the mirror symmetry with respect
to the plane is broken, for example, by an electric field or a substrate. For a graphene tube,
the Rashba coupling could be induced, for example, by an electric field transverse to the axis
of the tube. Kane and Mele found that, if $\lambda_R>2\sqrt{3}\lambda_{SO}$, then the graphene
would no longer be a QSH insulator \cite{Kane05}. For a smaller $\lambda_R$, even though the
graphene is still in the QSH phase with helical edge states, the electron spins in the edge
channels are no longer conserved. Therefore, the magnitude of the edge spin current is reduced.
It is interesting to investigate such a reduction in the spin transport. The ring geometry
proves to be most convenient since one does not have to worry about the disturbance from
external leads.

Fig.~\ref{rashba} shows the magnitude of the edge spin currents. When $\lambda_R=0$, the
magnitude is very close to $2$.  It gradually decreases to zero at the critical value of
$\lambda_R=2\sqrt{3}\lambda_{SO}$, beyond which the graphene is semi-metallic \cite{Kane05} and
there is no more edge spin current. Overall, the spin currents for spin-$x$ and spin-$y$
(calculated from semiclassical expressions similar to Eq.~(\ref{semi})) remain small for the
whole range of the Rashba coupling. One can see that, even if $\lambda_R$ is as large as
$\lambda_{SO}$, the spin current is still dominated by the spin-$z$ component. The slight
deviation from the ideal values of $2$ and $0$ on both ends is a result of the finite-size
effect and inter-edge coupling. Similar reduction of the spin in edge states, when the spin is
not conserved, has been observed in various spin-resolved measurements of Angle-Resolved
Photo-Emission Spectroscopy on three-dimensional topological insulators \cite{Hsieh09}.

\section{Conclusion}

By adding a spin-orbit interaction to a graphene ribbon, one can generate helical edge states
along armchair edges. When the ribbon is wrapped to form a ring that encloses a magnetic flux,
there are persistent charge and spin currents. We studied the persistent currents in rings with
different radii and widths. The bulk state contribution to the currents vanishes if the radius
of the ring is large, but the edge state contribution persists. For broad ribbons, the decay
length of the edge mode at the Dirac point is inversely proportional to $\lambda_{SO}$. The
persistent spin current for one helical edge is roughly $t/2\pi$. For narrow ribbons,
inter-edge coupling between the edge states opens the Dirac point (when $n_x\neq 3n+2$) and
reduces the persistent currents. For most of the parameters being studied, the Zeeman effect is
too small to have much influence. On the other hand, the Rashba coupling is found to
deteriorate the edge spin transport, which vanishes beyond a critical value when the graphene
is no longer a QSH insulator. Our work shows that the tubular QSH insulator driven by a
magnetic flux is an ideal configuration for theoretically exploring the edge transport.

Finally, a remark on the effect of electron-electron interaction, which is beyond the scope of
this work. In Ref.\cite{Soriano10}, it has been shown that the Hubbard type, short-ranged
interaction could induce edge magnetization. This is related to the following fact: the
honeycomb lattice is bi-partite, and the Hubbard interaction tends to induce antiferromagnetic
order. For a zigzag edge, since the lattice points along the edge are from the same
sub-lattice, their spins would point to the same direction. That is the main reason why a
zigzag edge is ferromagnetic. This mechanism will not work well for the armchair edge as it
contains lattice points from both sub-lattices. Therefore, magnetization induced by the Hubbard
interaction is less likely for the armchair edge.

\ack The authors would like to acknowledge the financial support from the National Science
Council of Taiwan.

\end{document}